\documentclass[a4paper,fleqn,usenatbib]{mnras}

\usepackage{newtxtext,newtxmath}

\usepackage[T1]{fontenc}
\usepackage{ae,aecompl}

\usepackage{graphicx}	
\usepackage{amsmath}	
\usepackage{amssymb}	

\title[King~1]{The open cluster King~1 in the second quadrant\footnote{Based on service observations made with the William Herschel Telescope operated on the island of La Palma by the Isaac Newton Group in the Spanish Observatorio del Roque de los Muchachos of the Instituto de Astrof\'{\i}sica de Canarias within the observing program SW2014b13.}}

\author[R. Carrera et al.]{
Ricardo Carrera,$^{1,2}$\thanks{E-mail: rcarrera@iac.es}
Loreto Rodr\'{\i}guez Espinosa,$^{1,2}$
Laia Casamiquela,$^{3}$
\newauthor
Lola Balaguer Nu\~nez,$^{3}$
Carme Jordi,$^{3}$
Carlos Allende Prieto,$^{1,2}$
\newauthor
and Peter B. Stetson$^{4}$
\\
$^{1}$Instituto de Astrof\'{\i}sica de Canarias, La Laguna E-3200, Tenerife, Spain\\
$^{2}$Departamento de Astrof\'{\i}sica, Universidad de La Laguna, La Laguna E-38205, Tenerife, Spain\\
$^{3}$Departament de F\'{\i}sica Qu\'antica i Astrof\'{\i}sica, Institut Ci\'encies Cosmos (ICCUB), Universitat de Barcelona, (IEEC-UB), Barcelona, Spain\\
$^{4}$National Research Council, 5071 West Saanich Road, Victoria, BC V9E 2E7, Canada 
}

\date{Accepted XXX. Received YYY; in original form ZZZ}

\pubyear{2015}

\begin{document}
\label{firstpage}
\pagerange{\pageref{firstpage}--\pageref{lastpage}}
\maketitle

\begin{abstract}
We analyse the poorly-studied open cluster King~1 in the second Galactic quadrant. From wide-field photometry we have studied the spatial distribution of this cluster. We determined that the centre of King~1 is located at  $\alpha_{2000}=00^{\rm h}22^{\rm m}$ and $\delta_{2000}=+64\degr23\arcmin$. By parameterizing the stellar density with a King profile we have obtained a central density of $\rho_{0}=6.5\pm0.2$ star arcmin$^{-2}$ and a core radius of $r_{\rm core}=1\farcm9\pm0\farcm2$. By comparing the observed color-magnitude diagram of King~1 with those of similar open clusters and with different sets of isochrones, we have estimated an age of $2.8\pm0.3$ Gyr, a distance modulus of $(m-M)_{\rm o}=10.6\pm0.1$ mag, and a reddening of $E(B-V)=0.80\pm0.05$ mag. To complete our analysis we acquired medium resolution spectra for 189 stars in the area of King~1. From their derived radial velocities we determined an average velocity $\left\langle V_r\right\rangle $=-53.1$\pm$3.1 km s$^{-1}$. From the strength of the infrared \mbox{Ca\,{\sc ii}} lines in red giants we have determined an average metallicity of $\left\langle [M/H]\right\rangle$=+0.07$\pm$0.08 dex. From spectral synthesis we have also estimated an $\alpha$-elements abundance of $\left\langle [\alpha/M]\right\rangle$=-0.10$\pm$0.08 dex.
\end{abstract}

\begin{keywords}
Stars: abundances -- Galaxy: disc -- Galaxy: open clusters and associations: individual: King~1
\end{keywords}



\section{Introduction}

King~1 is a poorly studied old open cluster located in the second Galactic quadrant in the direction of the 
Galactic anticentre at a Galactocentric distance of $R_{\rm gc}\sim$9.6 kpc and almost in the Galactic plane with a vertical distance of $z\sim$0.06 kpc \citep[e.g.][]{2002A&A...389..871D}. Moreover, it is located near the Perseus spiral arm. This cluster was discovered by \citet{1949BHarO.919...41K} during the examination of long-exposure photographs acquired with the 16-inch Metcalf refractor telescope located at the Oak Ridge Observatory (USA). To our knowledge, the first study of this cluster was performed by \citet{2004BASI...32..371L}, who acquired CCD photometry in Johnson \textit{UBV} Cousins \textit{RI} systems in the line of sight of King~1. The obtained color-magnitude diagram shows a broad main sequence with a clearly visible red clump. They derived a radius of 4\arcmin~ from its stellar surface density. They obtained an age of $1.6\pm0.4$ Gyr, a distance 
modulus of $(m-M)_{\rm o}=11.38$ mag, and a reddening of $E(B-V)=0.70\pm0.05$ mag, from isochrone fitting assuming solar metallicity. Using the same technique and \textit{BV} wide-field 
photometry, \citet{2007A&A...467.1065M} obtained an age of $\sim$4~Gyr, a 
distance modulus of $(m-M)_{\rm o}=10.17^{+0.32}_{-0.51}$ mag, and a reddening of 
$E(B-V)=0.76\pm0.09$ mag, also assuming solar metallicity. They determined a limiting radius of $r_{\rm lim}$=12\farcm3, defined as the radius where the cluster outskirts merge with the stellar background. In the same way, they derived a core radius of $r_{\rm core}$=2\farcm 1$\pm$0\farcm 1 defined as the distance between the centre and the point where the radial density profile becomes half of the central density. Shortly after, \citet{2008PASJ...60.1267H} derived an age of 
2.8 Gyr, a distance modulus of $(m-M)_{\rm o}=11.57$ and a reddening of $E(B-V)=0.62$ mag, 
when assuming solar metallicity from \textit{VI} photometry. Although \citet{2004BASI...32..371L} obtained an age more than 1 Gyr younger than that derived by \citet{2008PASJ...60.1267H}, the distance modulus and reddeding derived by these authors agree within 
the uncertainties. However, the distance modulus and age derived by 
\citet{2007A&A...467.1065M} are not compatible with those obtained by the 
other two studies.

To our knowledge, the first spectroscopic study in the area of King~1 was performed by \citet{2015A&A...578A..27C}. They obtained medium resolution spectra, with a resolution power of $\sim$8,500, in the near infrared \mbox{Ca\,{\sc ii}} triplet region, $\sim$8,500\AA, for ten stars selected in the region traditionally assigned to the red clump in the color-magnitude diagram of King~1. However, while other clusters studied by these authors outline a peaked distribution, King~1 did not show a clear peak in the radial velocity distribution.

The lack of agreement between the different photometric works that have studied King~1 and the fact that a clear peak in the radial velocity distribution was not found motivated us to perform a more detailed study of this cluster, which is presented in this paper.

The paper is organized as follows. The observational material is described in Section~\ref{sec2}. The determination of the centre of the cluster and its radial density profile are discussed in Section~\ref{sectpho}. The color-magnitude diagram of King~1 is presented in Section~\ref{sectcmd} where the age, distance modulus and reddening are estimated by comparing with similar clusters. In Section~\ref{sec31} the radial velocity determination is explained, and the obtained velocity distribution is discussed. The determinations of metallicity by comparison observed spectra with a  grid of synthetic spectra and from the strength of the infrared \mbox{Ca\,{\sc ii}} lines for giants are presented in Section~\ref{sec4}. Finally, the obtained results are discussed in Section~\ref{sectrends} in the context of the trends described by open clusters in the Galactic disc and the main conclusions of this work are given in Section~\ref{sec:conclusions}.

\section{Observational material and data reduction}\label{sec2}

\subsection{Imaging}\label{sec21}

\begin{figure}
	\includegraphics[width=\columnwidth]{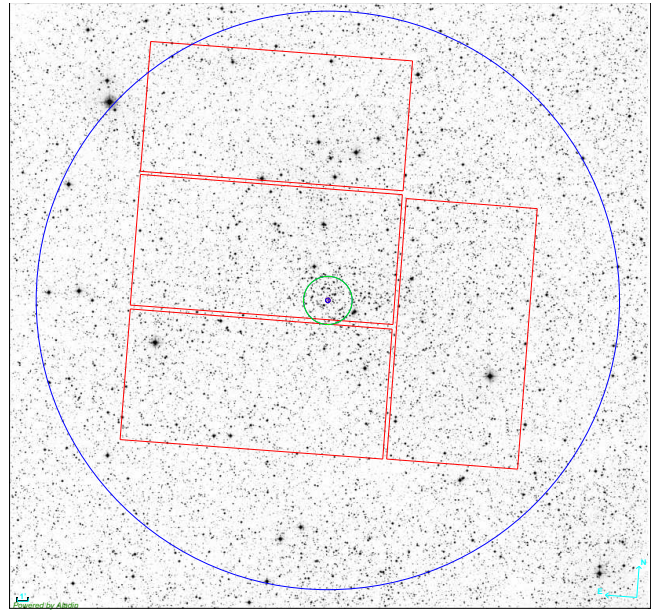}
    \caption{Digitized Sky Survey image of the King~1 region. The area covered by the WFC (red) and AF2 (blue) are overplotted. The green circle denotes the core radius (see text). North is up, and east is to the left.}
    \label{fig:wfcfov}
\end{figure}

Wide-field imaging of King~1 field was obtained in Johnson \textit{BV} and Sloan \textit{i} filters with the Wide Field Camera (WFC) installed at the prime focus of the 2.5 m Isaac Newton Telescope (INT) at the Roque de los Muchachos Observatory (La Palma, Spain). This instrument provides a total field of view of 34\farcm 2 $\times$ 34\farcm 2 (see Fig.~\ref{fig:wfcfov}). The observations were performed the nights of September 24 and 26, 2011 and August 14, 2015. A total of 4$\times$300 sec, 4$\times$600 s, and 4$\times$900 s were obtained in \textit{B}, \textit{V}, and \textit{i} bandpasses, respectively, together with 3$\times$30 s in each filter. Several standard fields and other open clusters were also observed these nights.

Bias subtraction, flat-field correction and removal of fringing in the \textit{i} band were performed using custom-made tasks developed by one of us (P.B. Stenson). The complete photometric catalogue is available at the extensive photometric database maintained by one of us \citep[P.B. Steson, see ][]{2000PASP..112..925S}\footnote{http://www.cadc-ccda.hia-iha.nrc-cnrc.gc.ca/en/community/STETSON/homogeneous/}. The photometric reduction was performed using \textit{DAOPHOT}/\textit{ALLFRAME} in the same way that other fields included in this data base \citep{1987PASP...99..191S,1994PASP..106..250S}.  The point spread function (PSF) was modeled for each individual image, sampling the whole area of each chip with a large number of stars. 

\begin{figure}
	\includegraphics[width=\columnwidth]{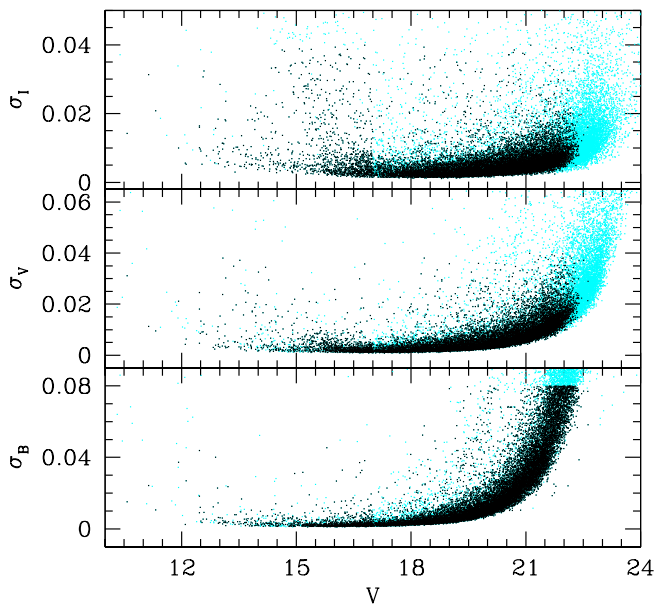}
    \caption{Final uncertainties for \textit{B} (bottom), \textit{V}, (middle), and \textit{I} (top) magnitudes as a function of \textit{V} magnitudes. Cyan points are all detected objects. Black points are selected stars after applying cuts in $\sigma_{B,V,I}$, $CHI$, and $SHARP$.}
    \label{fig:phosigma}
\end{figure}

After the \textit{ALLFRAME} run, the resulting magnitudes were aperture-corrected using \textit{DAOGROW} \citep[see][for details]{1990PASP..102..932S}. These magnitudes were transformed to the standard system following the procedure described in \citet{2000PASP..112..925S}. Basically, the instrumental magnitudes have been transformed to the standard system using nightly equations that include linear and quadratic color terms as well as linear extinction terms. Mean color coefficients have been determined and applied for all the nights of each observing run. Extinction coefficients and photometric zero points are determined on a night-by-night basis. Each transformation equation also includes linear terms in the x- and y-coordinates of the stellar image in the natural reference system of the CCD.  These correct any zero-point gradients which are probably due to uneven illumination of the detector during the flat-field exposures; these zero-point gradients typically amount to $\sim$0.01--0.03 mag from one edge of the
chip to the other, and are determined individually for each filter on each night.  Since several
hundred to several thousand standard stars are observed each night, all the transformation
coefficients are well defined. After the transformation equations for all nights have been determined, all the observations for each star are collected and transformed to the best possible magnitudes using a simultaneous least-squares optimization involving all available data for the star. Although we used the Sloan \textit{i} filter, we transformed the magnitudes to the widely used Cousins \textit{I} bandpass. Previous experience shows that Sloan \textit{i} filter can be transformed to Cousins \textit{I} magnitudes without adding additional uncertainty. In fact, in the case of King~1 observations, the median nightly root-mean-square residual of one observation of one star after transformation to Cousins \textit{I} system is 0.012 mag. This number includes both random photometric uncertainty and any systematic differences between the Cousins \textit{I} and Sloan \textit{i} photometric systems. The corresponding numbers for the uncertainties in the \textit{B} and \textit{V} magnitudes are 0.016 mag and 0.010 mag, respectively. The final uncertainties for each filter as a function of \textit{V} magnitudes are shown in Fig.~\ref{fig:phosigma}. Stars in the final list have been selected by rejecting those objects with $\sigma_B>$0.08, $\sigma_V>$0.04, and $\sigma_I>$0.04 to remove objects with large uncertainties, but also applying a cutoff in the value of $CHI>$1.3 and $SHARP<$-0.16 and $SHARP>$0.20. $SHARP$ and $CHI$ magnitudes are provided by \textit{ALLFRAME}. $SHARP$ is the difference between the width of the observed object and that of the PSF model. Therefore, it provides information about the object flattening. $CHI$ provides information about the quality of the PSF fit. 

\begin{figure}
\includegraphics[width=\columnwidth]{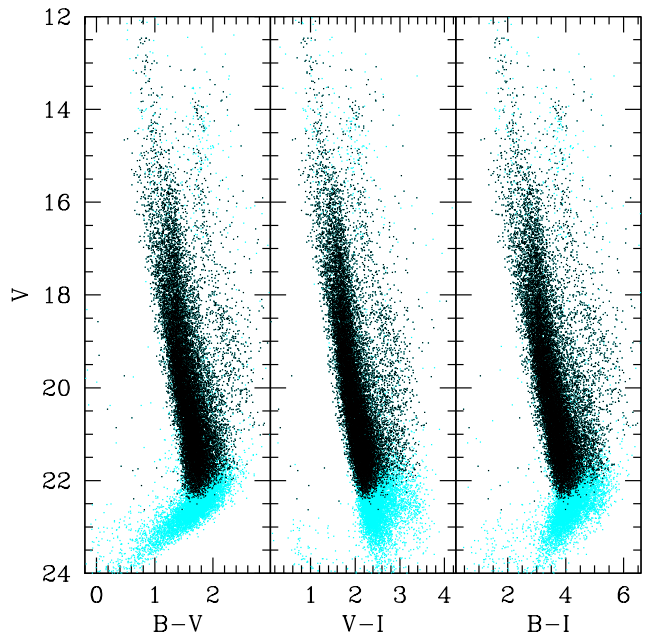}
\caption{The $B-V$ vs. $V$ (left), $V-I$ vs. $V$ (central), and $B-I$ vs. $V$ (right)  obtained color-magnitude diagrams. Cyan points are all observed stars. Black points denoted selected stars after applying rejections as a function of $\sigma_{B,V,I}$, $CHI$, and $SHARP$.}
\label{fig:dcm}
\end{figure}

The derived color-magnitude diagrams are shown in Fig.~\ref{fig:dcm}. A clear sequence is not observed in any of them and the diagrams appear dominated by foreground contamination. However, it is clear a clump of stars at 14$<V<$14.5 and 1.6$<(B-V)<$2.0 that may be related to the red clump of King~1. We postpone a detailed discussion of the derived color-magnitude diagrams to the forthcoming sections.

\subsection{Medium-resolution Spectroscopy}\label{af2spectra}

\begin{figure}
	\includegraphics[width=\columnwidth]{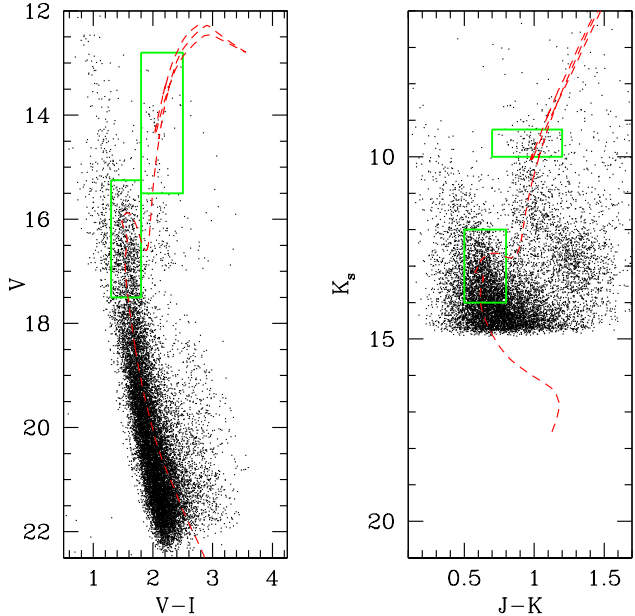}
    \caption{Location of the red clump and main sequence turn-off regions used to select the spectroscopic targets in the $V-I$ vs. $V$ (left) and $J-K_\mathrm{S}$ vs. $K_\mathrm{S}$ (right) color-magnitude diagrams. A 3 Gyr isochrone with solar metallity from BaSTI database \citep{2004ApJ...612..168P} has been over plotted as reference using the values derived in Section~\ref{sectcmd}}
    \label{fig:cmdsel}
\end{figure}

The color-magnitude diagram derived above has been used to select the targets for spectroscopic observations by defining two regions for the expected positions of the red clump and main sequence turn-off (left panel of Fig.~\ref{fig:cmdsel}). The observations were carried out the night of November 14, 2014 with  AutoFib2+WYFFOS (AF2)  multi-object spectrograph installed at the primary focus of the 4.2m William Herschel Telescope (WHT) at Roque de los Muchachos Observatory (La Palma, Spain).  It provides a field of view of about 50\arcmin\ of diameter. Since the area covered by AF2 is slightly larger than that covered by the WFC used to obtain the photometry (see Fig.~\ref{fig:wfcfov}), we extended the list of potential targets using the Two Micron All-Sky Survey \citep[2MASS;][]{2006AJ....131.1163S} by selecting stars in the same locations that those candidates selected previously in the $V-I$ vs $V$ color-magnitude diagram (right panel of Fig.~\ref{fig:cmdsel}). We used the R1200R grating centered at 8,500\,\AA, and the RG630 order-blocking filter. The RED+4 detector with 2$\times$2 binning yielding a dispersion of 0.8\, \AA~pix$^{-1}$ and a spectral resolution of R$\sim$8,000. Two different fiber configurations were observed in the King~1 field of view obtaining spectra for a total of 189 stars listed in Table~\ref{starsample}. For each configuration we acquired three expositions of 900 sec each.

Data reduction was performed using standard \textit{IRAF}\footnote{The Image Reduction and Analysis Facility, \textit{IRAF}, is distributed by the National Optical Astronomy Observatories, which are operated by the Association of Universities for Research in Astronomy, Inc., under cooperative agreement with the National Science Foundation.} routines. First, cosmic rays were removed from the images using the \textit{IRAF} Laplacian edge-detection routine \citep{2001PASP..113.1420V}. All images were then bias- and overscan-subtracted, and trimmed using {\sl ccdproc}. The spectra were extracted, flat-field corrected, and calibrated in wavelength using {\sl dofibers}, a task developed specifically to reduce data acquired with multi-fiber spectrographs. The lamp flats acquired at the beginning of each set-up were used to define and trace each aperture. Arcs, obtained before and after each configuration, were used for wavelength calibration.

An updated version of the procedure described in \citet{2008AJ....135..836C} and \citet{2011AJ....142...61C} was followed for sky subtraction. For each fiber in each individual exposure the nearest ten fibers placed on the sky were identified and combined to obtain a sky spectrum. By selecting only the nearest sky fibers, sky subtraction is improved in the sense that the residuals of sky lines are minimized. The resulting sky spectrum and the object spectrum for each fiber are separated into two components: continuum and line. To obtain the continuum of both sky and object lines we used a nonlinear median filter with 3-sigma clipping. The line spectrum is obtained by subtracting the continuum. The sky-line component is cross-correlated with the object-line one to put both in the same wavelength scale. This also provides an additional check of the wavelength calibration. The obtained offsets are very small, typically $\sim$0.01 pixels. After updating the wavelength calibration, the sky- and object-line components are compared to search for the scale factor that minimizes the sky line residuals over the whole spectral region. In practice, this optimum scaling factor is the value that minimizes the sum of the absolute differences between the object-line and the sky-line multiplied by the scale factor, known as L1 norm. The object-continuum component is wavelength updated and added back to the sky-subtracted object-line spectrum. Finally, the sky continuum is subtracted assuming that the scale factor is the same as for the sky-line component. 

\begin{table*}
 \centering
\caption{Stars observed with AF2. The full version of this table is available in the online journal and at CDS. $J$, $H$, and $K_\mathrm{S}$ magnitudes are from 2MASS.\label{starsample}} 
 \begin{tabular}{@{}lccccccccc@{}}
\hline
ID & $\alpha_{2000}$ & $\delta_{2000}$ & $V$ & $B$ & $I$ & $J$ & $H$ & $K_\mathrm{S}$ & $V_{r}$ \\
   &   (deg)         &      (deg)      &   (mag) & (mag)   & (mag)   &   (mag)& (mag)   &   (mag)  &   (km s$^{-1}$)\\
   \hline
2M00220155+6450572 & 5.506468 & +64.849243 & 99.999 & 99.999 & 99.999 & 13.285 & 12.850 & 12.715 & -29.8$\pm$3.2\\
2M00215836+6430433 & 5.493180 & +64.512039 & 16.322 & 17.598 & 14.758 & 13.639 & 13.198 & 13.034 &-104.4$\pm$5.6\\
2M00211378+6451046 & 5.307454 & +64.851280 & 99.999 & 99.999 & 99.999 & 12.953 & 12.526 & 12.347 & -10.4$\pm$2.5\\
2M00214805+6429184 & 5.450245 & +64.488472 & 16.852 & 18.001 & 15.410 & 14.326 & 13.959 & 13.714 & -50.8$\pm$3.5\\
2M00212009+6437258 & 5.333720 & +64.623840 & 16.110 & 17.281 & 14.705 & 13.525 & 13.106 & 12.939 & -10.8$\pm$3.5\\
\hline
\end{tabular}
\end{table*}

\section{Centre determination and radial density profiles}\label{sectpho}

We have calculated the cluster centre in $BVI$ and 2MASS catalogs independently by determining the maximum stellar density. Owing to the large foreground contamination in the line of sight of King~1, we have restricted our analysis to stars located in the expected positions of the cluster turn-off (see Fig.~\ref{fig:cmdsel}). By restricting our analysis to these areas we improved the completeness of our sample. The faintest magnitude used in our $BVI$ catalog, $V$=15.5, is about 7 magnitudes brighter than the magnitude limit of the sample $V\sim$22.5 mag. For 2MASS we used only stars brighter than $K_s$=14, which correspond to 99 \% completeness of 2MASS sample \citep{2006AJ....131.1163S}.

\begin{figure}
	\includegraphics[width=\columnwidth]{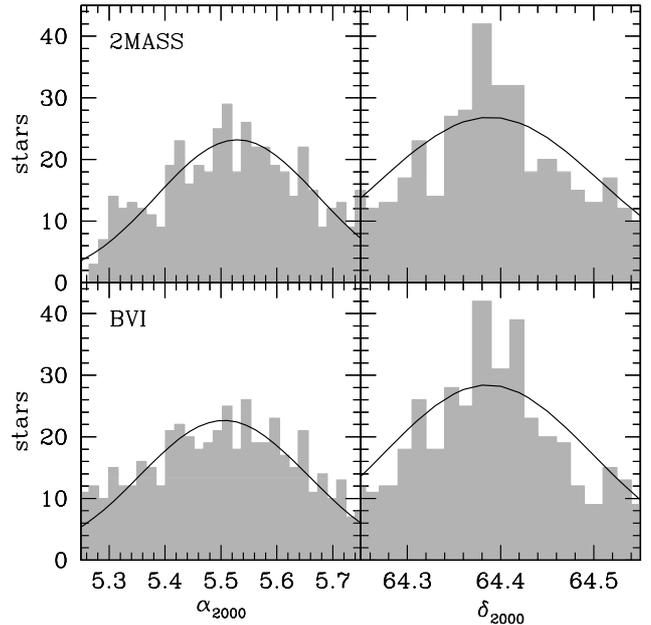}
    \caption{Stellar counts in a box of $\pm$15\arcmin~ around the first estimation of the cluster centre for 2MASS (top) and $BVI$ (bottom) samples. The Gaussians used to determine the centre in each case have also been plotted (see text for details).}
    \label{fig:centre}
\end{figure}

A first estimation of the centre of the cluster was obtained by fitting Gaussians to the histograms in right ascension, $\alpha$, and declination, $\delta$. We, then, constrain our analysis to objects located in a box of $\pm$15\arcmin~ around the first centre estimate in both right ascension and declination.  The histograms obtained for this sub-sample using a bin of $1\arcmin$ for $BVI$ and 2MASS are shown in the top and bottom panels of Fig.~\ref{fig:centre}, respectively. The centre of the cluster was obtained by fitting a Gaussian to each distribution: $\alpha_{2000}=5\fdg5054\pm19\arcsec$ and $\delta_{2000}=64\fdg3862\pm15\arcsec$, and $\alpha_{2000}=5\fdg5291\pm20\arcsec$ and $\delta_{2000}=+64\fdg3885\pm15\arcsec$ from $BVI$ and 2MASS samples, respectively. The centre of King~1, $\alpha_{2000}=00^{\rm h}22^{\rm m}$ and $\delta_{2000}=+64\degr23\arcmin$, was measured by averaging the results obtained for the $BVI$ and 2MASS samples. This result is, within the uncertainties, in agreement with the position determined by \citet{2007A&A...467.1065M} of $\alpha_{2000}=00^{\rm h}22^{\rm m}04^{\rm s}$ and $\delta_{2000}=+64\degr22\arcmin50\arcsec$, with an uncertainty of $\sim$1\arcmin.

\begin{figure}
	\includegraphics[width=\columnwidth]{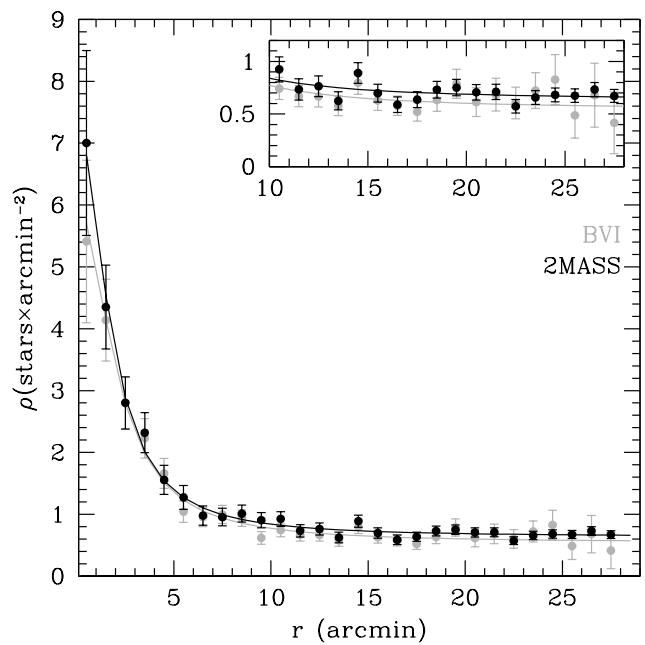}
    \caption{Stellar radial density profiles for $BVI$, grey, and 2MASS, black, samples. Radii outermost than 10\arcmin~ have been amplified in inset panel. Solid lines are the fits to the King profile (see text for details).}
    \label{fig:densityprofile}
\end{figure}

The radial density profile has also been computed in $BVI$ and 2MASS samples separately. Assuming the cluster centre determined above we have calculated the mean stellar surface density in concentric rings of 1\arcmin~ wide given by:

\begin{equation}
\rho_i=\dfrac{N_i}{\pi(R^2_{i+1}-R^2_{i})}
\end{equation}

\noindent where $N_i$ is the number of stars in the $i-th$ ring with inner and outer radius $R_i$ and $R_{i+1}$, respectively. The density uncertainty in each ring was estimated assuming Poisson statistics. Since the $BVI$ sample does not have a complete spatial coverage (see Fig.~\ref{fig:wfcfov}), we applied a scale factor that accounts for the ratio between the area covered and the total area of each ring. The obtained radial density profile for the $BVI$ and 2MASS samples are shown in Fig.~\ref{fig:densityprofile} in grey and black, respectively. The stellar surface density profiles  obtained have been characterized by fitting a King profile model \citep{1966AJ.....71...64K}:

\begin{equation}
\rho(r)=\rho_{bg}+\dfrac{\rho_{0}}{1+(\frac{r}{r_{\rm core}})^2}
\end{equation}

\noindent where $\rho_{bg}$ is the background density, $\rho_{0}$ is the central density, and $r_{\rm core}$ is the core radius, defined as the distance between the centre and the point where $\rho(r)=\frac{\rho_{0}}{2}$. 

We obtained a background density of $\rho_{bg}=0.54\pm0.04$ and $0.64\pm0.02$ stars arcmin$^{-2}$ for $BVI$ and 2MASS, respectively. The derived central densities are $\rho_{0}=6.4\pm1.1$ and $6.7\pm1.4$ star arcmin$^{-2}$ and the core radii are $r_{\rm core}=2\farcm1\pm0\farcm3$ and $1\farcm8\pm0\farcm3$ for $BVI$ and 2MASS, respectively. In all the cases the values obtained for the $BVI$ and 2MASS samples agree within the uncertainties. By averaging the results obtained in each bandpass we obtain a central density of $\rho_{0}=6.5\pm0.2$ star arcmin$^{-2}$ and a core radius of $r_{\rm core}=1\farcm9\pm0\farcm2$. These values are also similar, within the uncertainties, to those derived by \citet{2007A&A...467.1065M}. However, the background density derived by these authors, $\rho_{bg}=1.76\pm0.05$ stars arcmin$^{-2}$, is much larger than the obtained here. This can be explained by the fact that we have restricted our analysis to a region dominated by the King~1 population with low background contamination.

\section{King~1 color-magnitude diagram}\label{sectcmd}

\begin{figure}
	\includegraphics[width=\columnwidth]{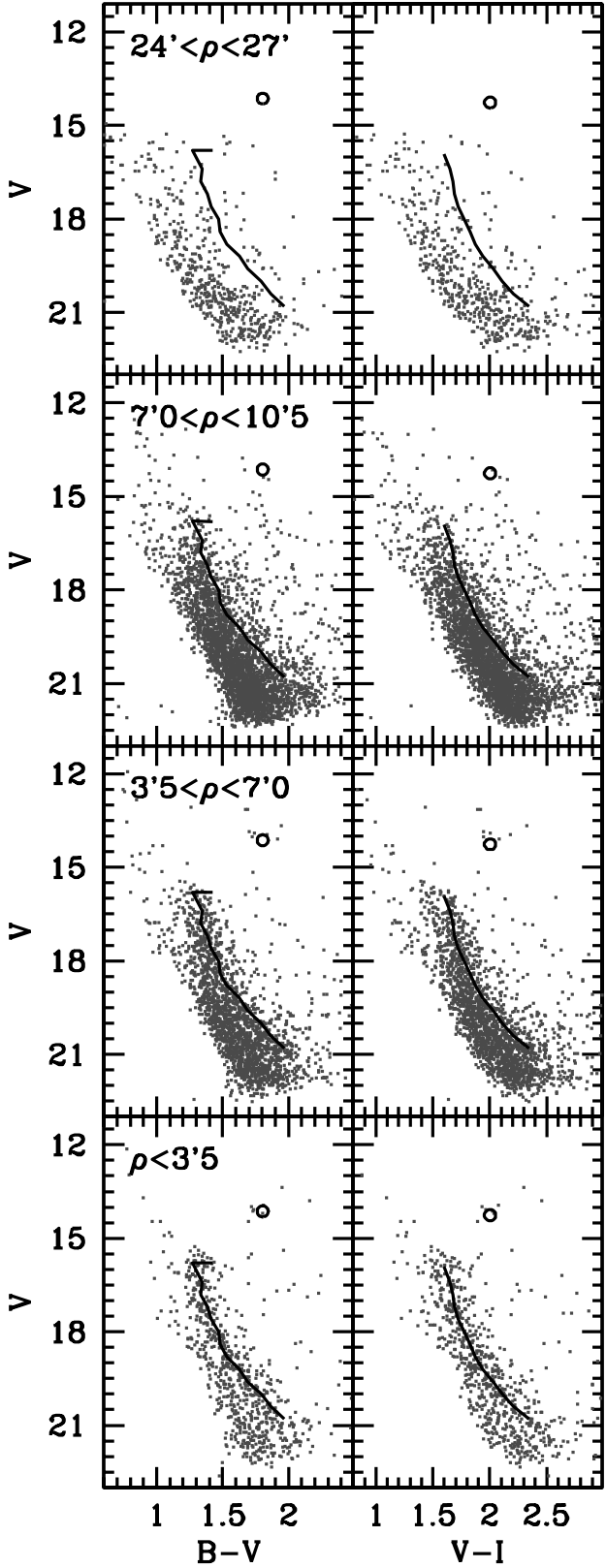}
    \caption{The $B-V$ vs. $V$ (left), $V-I$ vs. $V$ (right) color-magnitude diagrams at different distances from the cluster centre. Black lines and open circle show the cluster sequence (see text for details).  Top panel covers approximately the same area than the bottom one.}
    \label{fig:dcmrad}
\end{figure}

The obtained color-magnitude diagrams in the line of sight of King~1 are shown in Fig.~\ref{fig:dcm}, and appear dominated by foreground contamination. The color-magnitude diagrams of observed stars at different radii from the King~1 centre determined in previous section are plotted in Fig.~\ref{fig:dcmrad}. The King~1 sequence is clearly observed within the core radius (bottom panels of Fig.~\ref{fig:dcmrad}). The sequence is still visible until a radius of 7\arcmin, where the density has fallen to 20\% of the central value. Between 7\arcmin and 10$\farcm$5 the color-magnitude diagram is dominated by foreground contamination, although several stars are still observed in the position of the King~1 red clump. At a distance of 24\arcmin~ there are no sings of King~1 stars.

\begin{figure}
	\includegraphics[width=\columnwidth]{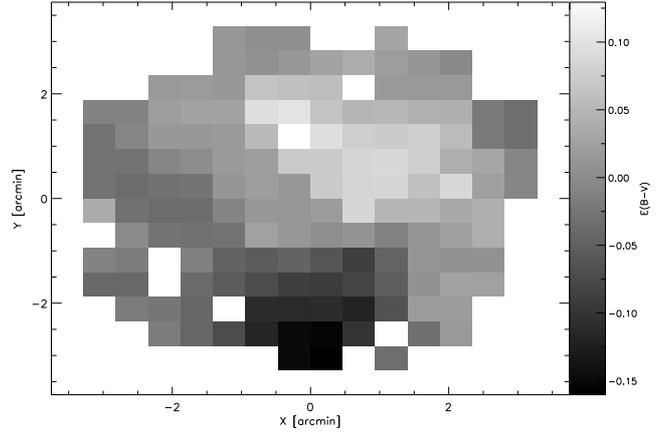}
    \caption{Map of differential reddening in the inner $3\farcm5$ of King~1. The grey levels correspond to the magnitude of the variation in local reddening, as indicated in the left panel.}
    \label{fig:reddeningmap}
\end{figure}

Before analyzing in detail the obtained color-magnitude diagram we have corrected for differential reddening. To do that we have applied the procedure described in detail by \citet{2012A&A...540A..16M} to the stars in the inner  $2\farcm5$.  Firstly, a photometric reference frame where the abscissa is parallel to the reddening line is defined. In this frame, we derived the fiducial main sequence ridge line. For each star, we selected the 25 nearest neighbors and evaluated the color distance from the main sequence ridge line for each of them. Derived offsets are applying in both color and magnitude to correct for local differential reddening. We do that only for stars in the main sequence. The obtained reddening map for the central  $3\farcm5$ of King~1 is shown in Fig.~\ref{fig:reddeningmap}.

\begin{figure}
	\includegraphics[width=\columnwidth]{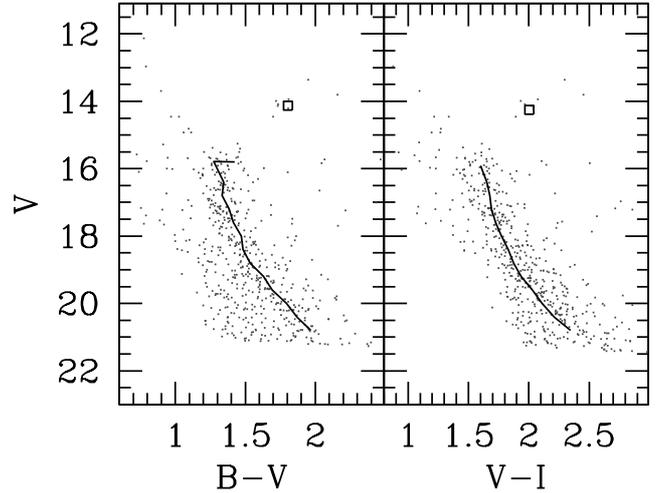}
    \caption{Differential reddening corrected color-magnitude diagrams in the inner $3\farcm5$. Black lines and open circle show the King~1  fiducial sequence (see text for details).}
    \label{fig:difred}
\end{figure}

The color-magnitude diagrams corrected for differential reddening are shown in Fig.~\ref{fig:difred}. From them we have derived the fiducial sequence of King~1. The range covered by the main sequence in magnitude was splitted in bins of 0.25 mag. In each bin the mean color was obtained using a sigma-clipping rejection and applying cuts in color to avoid background contamination. In the same way, the main sequence turn-off region was divided in bins of 0.1 mag. Again, the mean magnitude was obtained applying a sigma-clipping rejection. Finally, the location of the red clump was obtained by averaging the color and magnitude of stars in the red clump area. We have not tried to obtain a fiducial sequence for the red giant branch because it is not clearly observed in the color-magnitude diagram.

\begin{figure}
	\includegraphics[width=\columnwidth]{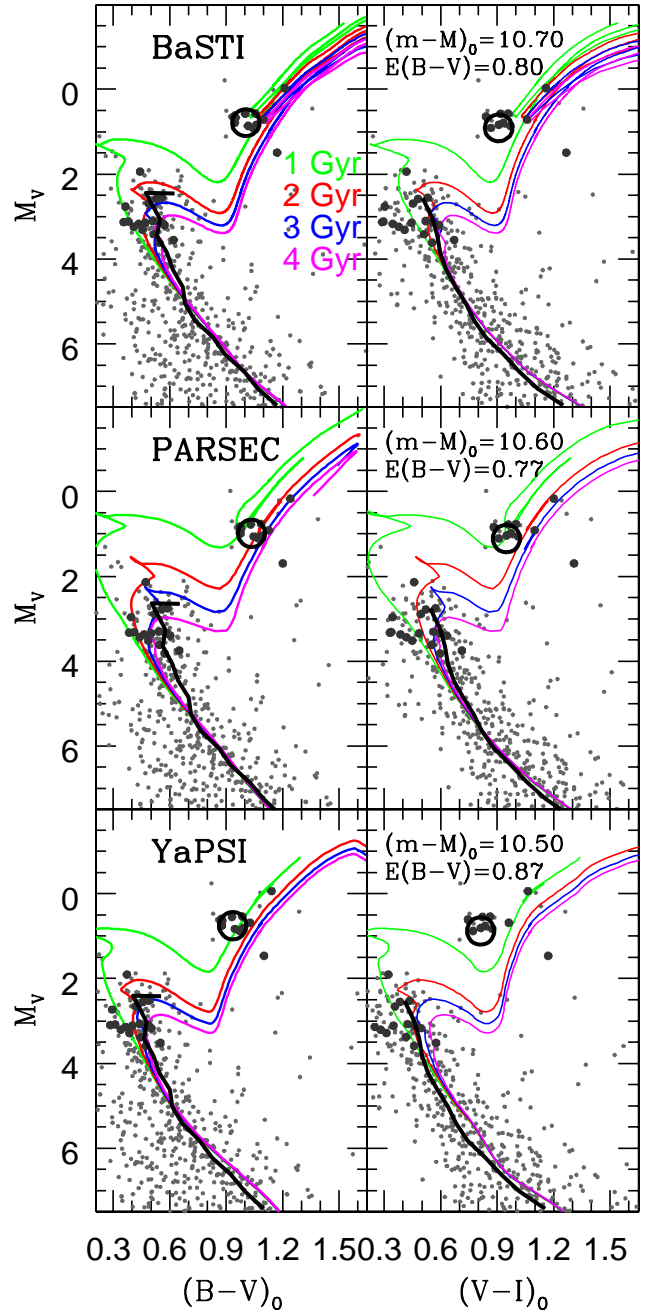}
    \caption{Isochrones overlay in the King~1 $(B-V)_{\rm o}$ (left) and  $(V-I)_{\rm o}$  (right) vs. $M_V$ color-magnitude diagrams. Dark grey points are the stars confirmed as King~1 members from their radial velocity (see Section~\ref{sec31}). Black line is the King~1 main sequence fiducial. Black open circle denotes the King~1 red clump position.}
    \label{fig:isocronas}
\end{figure}

The age, distance modulus and reddening are traditionally derived by over-imposing isochrones to the observed color-magnitude diagram. In this case we have used three widely used isochrones sets: a Bag of Stellar Tracks and Isochrones \citep[BaSTI;][]{2004ApJ...612..168P}; PAdova and TRieste Stellar Evolution Code \citep[PARSEC;][]{2012MNRAS.427..127B}; and Yale-Postdam Stellar Isochrones \citep[YaPSI;][]{2017ApJ...838..161S}. In all the cases we assumed a solar metallicity (see Section~\ref{sec4}). For BaSTI we used the canonical solar-scaled isochrones with Y=0.273 and $\eta$=0.2. In the case of PARSEC the webtool\footnote{http://stev.oapd.inaf.it/cgi-bin/cmd} was used to derive isochrones with Z= 0.0152. Note that this is the solar metallicity for PARSEC isochrones. Finally, for YaPSI we used the set with Y=0.28 and [Fe/H]=0.0. Isochrones with ages between 1 and 4 Gyr for each set have been overplotted to the King~1 color-magnitude diagram in Fig.~\ref{fig:isocronas}. The distance modulus, $(m-M)_{\rm o}$,  and a reddening, $E(B-V)$, have been tuned manually for each set of isochrones in order to better reproduce the observed diagram. Obtained values are listed in Table~\ref{agetable}. 
These parameters have been obtained trying to reproduce mainly the main sequence and turn-off. PARSEC and YaPSI 3 Gyr old isochrones reproduce well the observed  $(B-V)_{\rm o}$ color-magnitude digrams. In the case of BaSTI King~1 main sequence fiducial sequence lays in between 2 and 3 Gyr old isochrones. The distance modulus and reddening uncertainties have been estimated as the range where the isochrones still reproduce acceptable the observed diagram. For the age we have assumed a conservative uncertainty of 0.5 Gyr, half of the step between the isochrones used. The three sets reproduced relatively well the position of the red clump in the $(B-V)_{\rm o}$ diagram. However, all these models predict that its position in the $(V-I)_{\rm o}$ diagram is reddest than really observed. This discrepancy is larger in the case of YaPSI isochrones. This discrepancy has been already noticed by \citet{2013MNRAS.430..221A}. These authors showed that the difference between the observed and predicted color position of the red clump is larger for clusters with metallicity above solar. These authors attributed this discrepancy to the inability of past and present atmosphere models to reproduce real colours at high metallicity (i. e. low temperature).

\begin{table}
 \centering
\caption{Magnidutes derived from comparison with isochrones and other open clusters.\label{agetable}} 
 \begin{tabular}{@{}lccc@{}}
\hline
Set  & Age & (m-M)$_V$ & E(B-V) \\
 & (Gyr) & (mag) & (mag) \\ 
   \hline
BaSTI & 2.5$\pm$0.5 & 10.7$\pm$0.1 & 0.80$\pm$0.05 \\
PARSEC & 3.0$\pm$0.5 & 10.6$\pm$0.1 & 0.77$\pm$0.03 \\
YaPSI & 3.0$\pm$0.5 &  10.5$\pm$0.1 & 0.87$\pm$0.04  \\
OCs & 3.0$\pm$1.0 & 10.6$\pm$0.1 & 0.80$\pm$0.05  \\
\hline
Adopted & 2.8$\pm$0.3 & 10.6$\pm$0.1 & 0.80$\pm$0.05 \\
\hline
\end{tabular}
\end{table}

\begin{figure}
	\includegraphics[width=0.85\columnwidth]{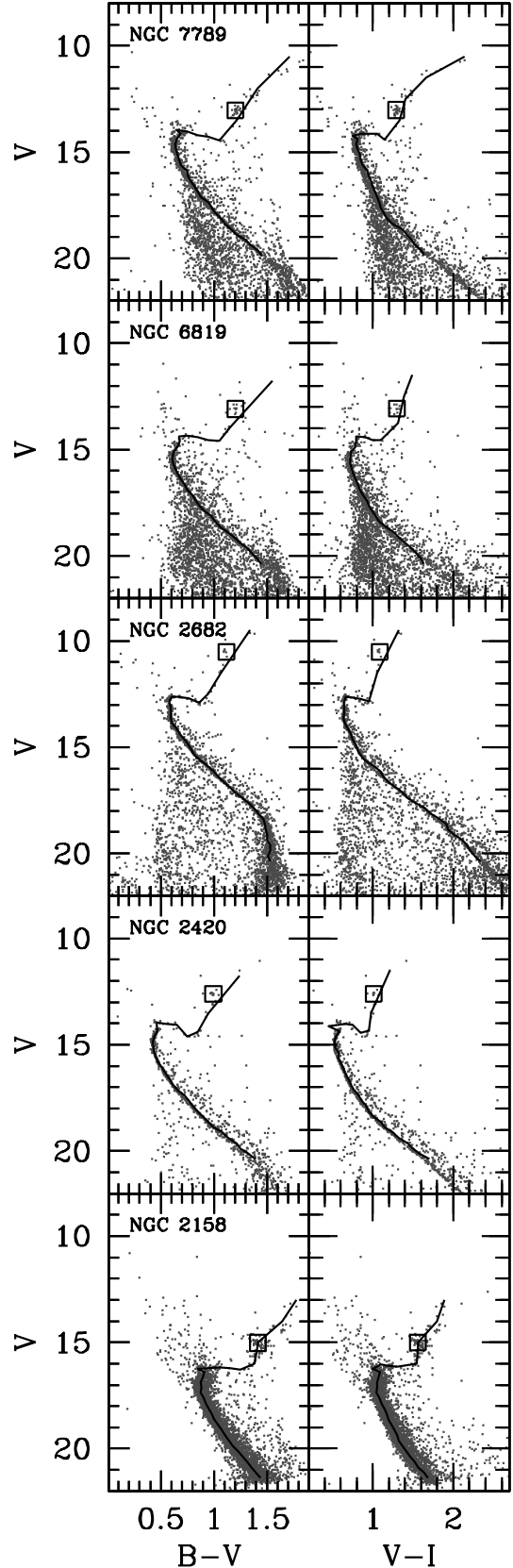}
    \caption{The $B-V$ vs. $V$ (left), $V-I$ vs. $V$ (right) color-magnitude diagrams of the open clusters used to estimate the  properties of King~1. As in Fig.~\ref{fig:dcmrad}, black squares are the fiducial sequences derived for each cluster.}
    \label{fig:dcmOCs}
\end{figure}

Complementary, we have compared the King~1 color-magnitude diagrams with those of other well studied open clusters in the literature. For this purpose we have chosen five open clusters with ages in the range of that expected for King~1 and also included in the Stetson's photometry data base \citep{2000PASP..112..925S}. These clusters are NGC~2158, NGC~2420, NGC~2682, NGC 6819, and NGC~7789. Their properties are summarized in Table~\ref{OCpropierties}. The fiducial sequences of these clusters have been obtained following the same procedure than in the case of King~1. In some of them the sequence parallel to the main sequence defined by binaries is clearly observed. Although our method is able to separate the main- and binary-sequences in order to compute the fiducial sequence we have tried to avoid any bias caused by the presence of binaries. The fiducial red giant branch sequences have been obtained following a similar procedure than in the case of the main sequence but using bins of 1 mag. Their color-magnitude diagrams and the derived fiducial sequences are shown in Fig.~\ref{fig:dcmOCs}.

\begin{table}
 \centering
\caption{Properties of the comparison open clusters.\label{OCpropierties}} 
 \begin{tabular}{@{}lcccc@{}}
\hline
Cluster  & Age & (m-M)$_V$ & E(B-V) & [Fe/H] \\
 & (Gyr) & (mag) & (mag) & (dex) \\ 
   \hline
NGC~2158 & 1.9$\pm$0.2$^1$ & 14.28$\pm$0.06$^1$ & 0.42$\pm$0.09$^1$ & -0.01$\pm$0.05$^2$ \\
NGC~2420 & 2.0$\pm$02$^3$ & 11.88$\pm$0.27$^4$ & 0.04$\pm$0.03$^4$ & -0.10$\pm$0.04$^5$ \\
NGC~2682 & 4.2$\pm$0.2$^6$ &  9.67$\pm$0.11$^4$ & 0.04$\pm$0.02$^4$ & +0.04$\pm$0.03$^5$ \\
NGC~6819 & 2.3$\pm$0.1$^7$ & 12.38$\pm$0.04$^7$ & 0.13$\pm$0.02$^7$ & +0.09$\pm$0.03$^5$ \\
NGC~7789 & 1.6$\pm$0.5$^4$ & 12.23$\pm$0.20$^4$ & 0.27$\pm$0.04$^4$ & +0.06$\pm$0.05$^5$\\
\hline
\end{tabular}
\begin{minipage}{80mm}
References: (1) \citet{2010ApJ...708L..32B}; (2) \citet{2011AJ....142...59J}; (3)  \citet{2000AJ....120.1384V}
(4) \citet{2010A&A...511A..56P}; (5) \citet{occaso2017}; (6) \citet{2016ApJ...823...16B}; (7) \citet{2016AJ....151...66B}
.
\end{minipage}
\end{table}

\begin{figure}
	\includegraphics[width=\columnwidth]{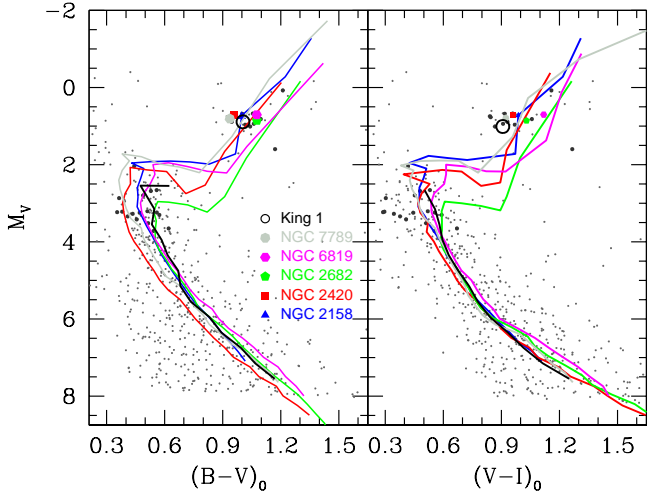}
    \caption{As Fig~\ref{fig:isocronas} with the comparison with the fiducial sequences of other open clusters.}
    \label{fig:fit}
\end{figure}

The fiducial sequences of the comparison clusters have been over-plotted in the color-magnitude diagram of King~1 in Fig.~\ref{fig:fit} using the distance modulus and reddening listed in Table~\ref{OCpropierties}. We have tried to constrain at the same time the distance modulus, reddening and age of King~1 by matching at the same time the main sequence and the red clump positions. The best solution is shown in Fig.~\ref{fig:fit}. Using a distance modulus of $(m-M)_{\rm o}$=10.6 mag and a reddening of $E(B-V)$=0.80 mag we are able to locate the King~1 red clump in the same position that the comparison clusters. The main sequence of King~1 follows a similar trend than those of NGC~2682 and NGC~6819. This implies that the three systems may have a similar metallicity. Though the turn-off is not well delimited, as commented before, it seems that it lies in between those of NGC~2682 and NGC~6819. Therefore, we assigned an age of 3 Gyr to King~1 with an uncertainty of 1 Gyr. For distance modulus and reddening we assumed uncertainties of 0.1 and 0.05 mag, respectively. Our age estimation and the comparison shown in Fig.~\ref{fig:fit} will not change significantly if these quantities varied within these ranges. Outside them the match gets worse.

Obtained values from isochrones and from comparison with other open clusters are in very good agreement within the uncertainties. For these reason we have adopted for King~1 the weighted mean of the values derived from the three isochrone sets and the open clusters comparison. These values are listed in the last row of Table~\ref{agetable}. The age derived here is similar, within the uncertainties, to the values derived by \citet{2007A&A...467.1065M} and \citet{2008PASJ...60.1267H} but older than the estimation of $1.6\pm0.6$ Gyr obtained by \citet{2004BASI...32..371L}. Our reddening is similar, within the uncertainties, to the value derived by \citet{2007A&A...467.1065M} but larger than those provided by \citet{2004BASI...32..371L} and \citet{2008PASJ...60.1267H}. Finally, the distance modulus is in between other values available in the literature. \citet{2004BASI...32..371L} and \citet{2008PASJ...60.1267H} obtained $(m-M)_{\rm o}$=11.38 and 11.57, respectively, whereas \citet{2007A&A...467.1065M} obtained $(m-M)_{\rm o}=10.17^{+0.32}_{-0.51}$. All these values have been derived by comparing the observed color-magnitude diagrams with isochrones.

\section{Radial velocity distribution}\label{sec31}

The radial velocities of the stars observed with AF2 (Section~\ref{af2spectra}) were calculated using the classical cross-correlation method \citep[e.g.][]{1979AJ.....84.1511T} by comparing the observed spectra with a grid of synthetic spectra of different features \citep{2007AJ....134.1843A}. In our case, the grid of synthetic spectra has been computed with ASS$\varepsilon$T \citep{2008ApJ...680..764K,2009AIPC.1171...73K} using ATLAS9 stellar atmosphere models \citep{2003IAUS..210P.A20C}. This grid has five dimensions: metallicity, [M/H]; $\alpha$-elements abundance (O, Mg, Si, S, Ca, and Ti), [$\alpha$/M]; micro-turbulence, $\xi$; effective temperature, $T_{\rm eff}$; and surface gravity, log $g$ with the ranges and steps listed in Table~\ref{griddimension}. The abundances of other elements were fixed to the Solar values \citep{2005ASPC..336...25A}. In total, this grid contains 136,125 synthetic spectra.

\begin{table}
 \centering
\caption{Range of each dimension in the synthetic grid used to the radial velocity determination.\label{griddimension}} 
 \begin{tabular}{@{}lccc@{}}
\hline
Dimension  & First & Last & Step \\
   \hline
$[$M/H$]$ (dex)& -5.0 & +1.0  & 0.25\\
$[\alpha$/M$]$ (dex)& -1.0  & +1.0   &  0.25\\
log $\xi$ (dex) & -0.3  & 1.2  &  0.3\\
$T_{\rm eff}$ (K) & 3500  & 6000  & 250\\
log $g$ (dex) & 0.0  &  5  & 0.5\\
\hline
\end{tabular}
\end{table}

\begin{figure}
	\includegraphics[width=\columnwidth]{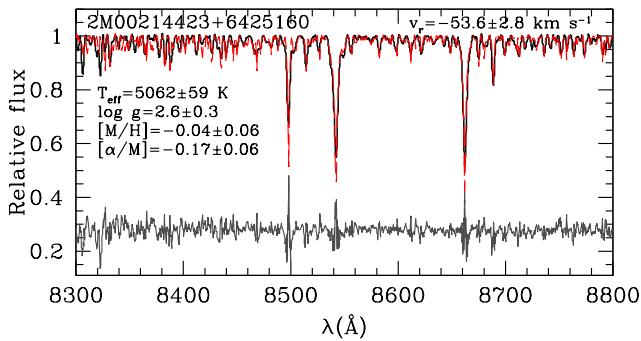}
    \caption{Normalized observed spectrum of star 2M00214423+6425160 (black), the best fit (red) and the residuals of the fit (grey). }
    \label{fig:ajuste}
\end{figure}

\begin{figure}
	\includegraphics[width=\columnwidth]{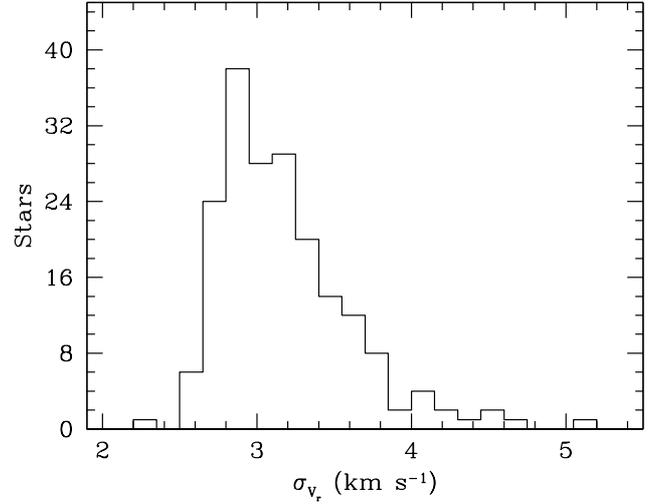}
    \caption{Histogram of the $\sigma_{V_{r}}$ of the stars observed with AF2. The median of the distribution is 3.1 km s$^{-1}$ with a standard deviation} of 0.4 km s$^{-1}$. 
    \label{fig:vrsig}
\end{figure}

In brief, radial velocities are determined using the following procedure. (i) Each object spectrum is cross-correlated with a reference synthetic spectrum to obtain an initial shift. In this case we selected one with the Arcturus parameters: [M/H]=-0.5 dex; [$\alpha$/H]=+0.0 dex; $\xi$=1.5 km s$^{-1}$; T$_{\rm eff}$=4,500 k; and log $g$=2.0 dex. (ii) After applying this initial shift, the observed spectrum is compared with the whole grid in order to identify the model parameters that best reproduces it through a $\chi^2$ minimization using FER\reflectbox{R}E\footnote{Available at https://github.com/callendeprieto/ferre} \citep{2006ApJ...636..804A}. An example of the obtained fits is shown in Fig.~\ref{fig:ajuste}. (iii) The best fit synthetic spectrum is cross-correlated again with the observed spectrum in order to refine the shift between both. Of course, Arcturus is not the ideal template for main sequence stars. However, after the first initial determination of the shift with Arcturus, our procedure converges by itself to appropriate templates for these objects. Therefore, the final radial velocities for these objects are derived using the best templates for them. The heliocentric radial velocities derived for each star are listed in last column of Table~\ref{starsample}. The distribution of the uncertainties of the derived radial velocities obtained from the width of the correlation peak is shown in Fig.~\ref{fig:vrsig}.

\begin{figure}
	\includegraphics[width=\columnwidth]{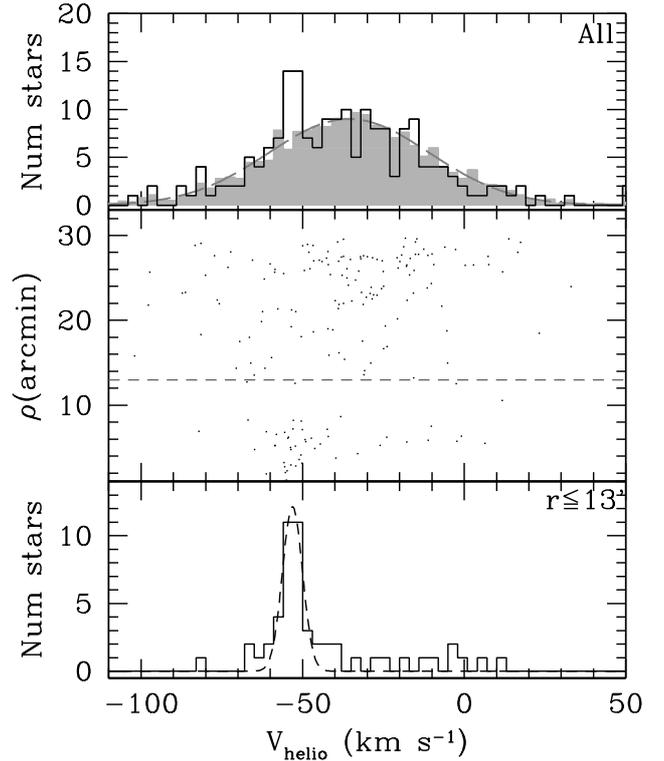}
    \caption{Top: Radial velocity distribution of observed stars (black histogram). The velocity distribution predicted by the Besan\c{c}on Galaxy Model at the location of King~1 has been over-plotted as comparison (shadow grey histogram).  The dashed line is the Gaussian fit to the Besan\c{c}on Galaxy Model velocity distribution prediction. Middle: Radial velocity as a function of distance to the King~1 centre. Bottom: Radial velocity distribution of observed stars in the inner 13\arcmin (dashed line in middle panel). As before, dashed line is the Gaussian fit.}
    \label{fig:distvr}
\end{figure}

The velocity distribution of the observed stars is shown as black histogram in top panel of Fig.~\ref{fig:distvr}. 
This distribution may be affected by differential gravitational red-shifts  and convective blue-shifts, since we are combining stars on the upper main sequence and on the red giant branch. The expected impact of these corrections on the derived radial velocities is under  1 km s$^{-1}$ \citep[see, e. g.][]{2002ApJ...566L..93A,2011A&A...526A.127P}. This value is a third of our typical uncertainties, and therefore we have made no attempt to correct it.

To better understand the obtained velocity distribution we have compared it with a model prediction. Using the Besan\c{c}on Galaxy Model\footnote{Available at http://model.obs-besancon.fr/} \citep{2003A&A...409..523R} we have computed a model of the expected stellar population at the position of King~1. We considered observational errors in magnitudes using similar uncertainties to those discussed in Sect~\ref{sec21} for the photometry . For the radial velocities we assumed a typical uncertainty of 3 km s$^{-1}$, similar to that for our sample (see Fig.~\ref{fig:vrsig}). We restricted our comparison to stars located in the same regions of the color-magnitude diagram as those observed here. The obtained velocity distribution has been over-plotted in the top panel of Fig.~\ref{fig:distvr} as shadow grey histogram. It has been scaled to the number of the observed stars. Except for the peak between -60 and -50 km s$^{-1}$ both distributions match reasonable well. We have fitted a Gaussian to the model velocity distribution (long dashed grey line) obtaining a mean value of -35.7$\pm$0.4 km s$^{-1}$ with a dispersion of 25.0$\pm$0.4 km s$^{-1}$.

As commented before, \citet{2015A&A...578A..27C} derived radial velocities for ten stars in the line of sight of King~1. The obtained  velocity distribution did not show a clear peak in contrast to the sample for other clusters studied in the same paper. From these stars they derived an average velocity of $\langle V_{r}\rangle=-38.4$ km s$^{-1}$ with a dispersion of $\pm11.6$  km s$^{-1}$ although the authors state that they cannot confirm that they detected real King~1 members.  Since this value is quite similar to the one derived by \citet{2015A&A...578A..27C} this could imply that their sample was dominated by field contamination. Our AF2 sample has five stars in common with the \citet{2015A&A...578A..27C}. On average, the radial velocities derived here are 15 km s$^{-1}$ larger than those derived by \citet{2015A&A...578A..27C}. A problem with the wavelength calibration of the King~1 spectra analysed by \citet{2015A&A...578A..27C} may explain this discrepancy.

To confirm that the peak between -60 and -50 km s$^{-1}$ may correspond to King~1 stars, we have plotted in the bottom panel of Fig.~\ref{fig:distvr} the radial velocity distribution of stars located in the inner 13\arcmin, where the density has fallen to about a 10\% of the central one. In this case, the peak is very prominent with only a few stars outside it. We have fitted a Gaussian to this peak, dashed black line. We obtained an average velocity of 
-53.1$\pm$0.2 km s$^{-1}$ with a dispersion of 3.1$\pm$0.2 km s$^{-1}$.

\section{Metallicity determination}\label{sec4}

\begin{figure}
	\includegraphics[width=\columnwidth]{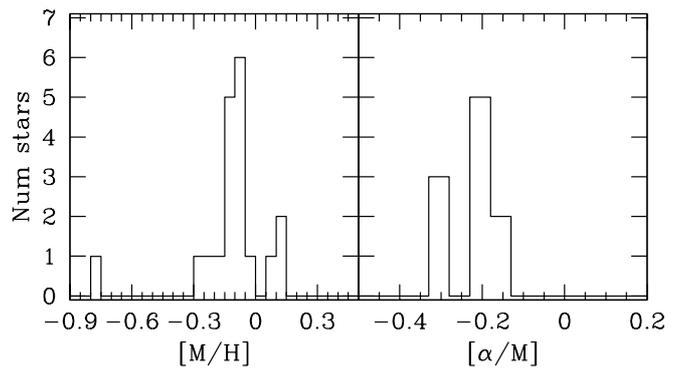}
    \caption{[M/H]$_{fit}$ (left) and [$\alpha$/M]$_{fit}$ (right) distributions of the stars in the innermost 13\arcmin\ and with radial velocities in the range $-53.1\pm2\times3.1$ km s$^{-1}$.}
    \label{fig:synthabu}
\end{figure}

We determine metallicities by two methods. The comparison between the observed spectra and the grid of synthetic spectra performed in the determination of radial velocities provides additional information about the chemical composition of the observed stars. In order to constrain the metallicity of King~1 we have restricted our analysis to those stars within the innermost 13\arcmin\@ and with radial velocities within  $\pm$2$\times$3.1  km s$^{-1}$ around the cluster average of -53.1 km s$^{-1}$. A total of 28 objects, 11 giant and 17 main sequence stars, meet these criteria. Moreover, we excluded of the analysis 9 main sequence stars with signal-to-noise ratios lower than 20 for which FER\reflectbox{R}E  does not provide reliable fits for the whole spectral range. The [M/H]$_{fit}$ and [$\alpha$/M]$_{fit}$ distributions for the remaining 19 objects are shown in the left and right panels of Fig.~\ref{fig:synthabu}, respectively. These values are listed in Table~\ref{tablecat}. Both distributions show clear peaks. We have computed the weighted mean of each of them obtaining $\left\langle [M/H]_{fit}\right\rangle $=-0.06 dex with $\sigma$=0.05 dex with an average uncertainty of $\pm$0.20 dex obtained as the median of the individual uncertainties for each star. For the $\alpha$-elements abundance we obtained $\left\langle [\alpha/M]\right\rangle $=-0.22 dex and $\sigma$=0.08 dex with an average uncertainty of $\pm$0.25 dex. 

It is necessary to check if these results are affected by zero points due to systemic errors. For this purpose we have analysed two well know stars: Arcturus and $\mu$ Leo. Since these stars have not been observed with AF2, we have smoothed the high-resolution high signal-to-noise spectra obtained with HERMES at the Mercator telescope (La Palma, Spain) in the framework of the Open Clusters Chemical Abundance from Spanish Observatories project \citep[OCCASO;][]{2016MNRAS.458.3150C}  to match the AF2 resolution. Moreover, we have degraded the signal-to-noise ratio of these spectra to $\sim$20 to match those of the spectra used in this work. Obtained values applying the same procedure than in the case of King~1 are listed in Table~\ref{synthcal} together with the reference ones obtained from \citet{occaso2017}. The values obtained with the spectral synthesis for the two comparison stars are slightly lower than the reference ones although they can be considered as consistent within the uncertainties. The value of $[$M/H$]$ obtained for $\mu$ Leo is slightly higher. In the case of the $\alpha$-elements abundance the values obtained are, on average, 0.12 dex lower than the reference ones. Therefore, the value obtained above for King~1 is likely underestimated by about 0.12 dex. Correcting for this offset, we arrive at an average $\left\langle [\alpha/M]\right\rangle $=-0.10 dex.

\begin{table}
 \centering
\caption{Obtained values for Arcturus and $\mu$ Leo and the reference ones. \label{synthcal}} 
 \begin{tabular}{@{}lcccc@{}}
\hline
  & \multicolumn{2}{c}{Arcturus} & \multicolumn{2}{c}{$\mu$ Leo}\\
  & fit & Ref & fit & Ref \\ 
   \hline
$[$M/H$]$ (dex)& -0.65$\pm$0.02 & -0.55$\pm$0.03 & +0.35$\pm$0.02 & +0.30$\pm$0.08\\
$[\alpha$/M$]$ (dex)& +0.16$\pm$0.02 & +0.25$\pm$0.05 & -0.05$\pm$0.02 & +0.08$\pm$0.06\\
$T_{\rm eff}$ (K) & 4174$\pm$30 & 4257$\pm$50 & 4398$\pm$30 & 4471$\pm$40\\
log $g$ (dex) & 1.09$\pm$0.02 & 1.6$\pm$0.15 & 2.29$\pm$0.02 & 2.3$\pm$0.1\\
\hline
\end{tabular}
\end{table}

Additionally, the strength of the infrared \mbox{Ca\,{\sc ii}} triplet lines provides an alternative method to determine metallicities in red giants using the relations available in the literature \citep[e.g.][]{2013MNRAS.434.1681C}. We have determined the strength of these lines in the 11 giants that meet the criteria described above using the procedure described by \citet{2007AJ....134.1298C}. In 
brief, the equivalent width of each line is determined as the area between the 
line profile and the continuum. The line profile is determined by fitting a Gaussian plus 
a Lorentzian, which 
provides the best fit to the line core and wings. Although the spectrum was 
previously 
normalized, the continuum level is recalculated by performing a linear fit
to the mean values of several continuum 
bandpasses defined for this purpose. The bandpasses used to fit the
line profile and to determine the continuum position are those described by
\citet{2001MNRAS.326..959C}. The equivalent widths of 
each line and their uncertainties, which were determined for each star, are listed in 
Table~\ref{tablecat}. Finally, the Calcium triplet index, denoted $\Sigma$Ca, was obtained as 
the sum of the equivalent widths of the three lines. 

\begin{table*}
 \centering
\caption{Individual [M/H] and [$\alpha$/M] values obtained from FER\reflectbox{R}E and the strength of the infrared \mbox{Ca\,{\sc ii}} lines and derived [M/H] from them in each bandpass. See text for details.\label{tablecat}} 
 \begin{tabular}{@{}lccccccccccc@{}}
\hline
ID  & [M/H]$_{fit}$ & [$\alpha$/M]$_{fit}$ & EW$_{8498}$ & EW$_{8442}$ & EW$_{8662}$ & [M/H]$_V$ & [M/H]$_I$ & [M/H]$_{K_s}$ \\
    &     (dex)  &	(dex) &  (\AA)      &     (\AA)   &    (\AA)    &   (dex)    &     (dex)  &	(dex)   \\
   \hline
2M00215876+6423200 & -0.30$\pm$0.26 &  & 1.63$\pm$0.03 & 3.47$\pm$0.02 & 2.79$\pm$0.03 & +0.04$\pm$0.06 & +0.09$\pm$0.04 & +0.04$\pm$0.04\\
2M00213494+6423181 & -0.12$\pm$0.20 & -0.22$\pm$0.07 & 1.56$\pm$0.03 & 3.53$\pm$0.02 & 2.85$\pm$0.02 & +0.01$\pm$0.06 & +0.10$\pm$0.04 & +0.01$\pm$0.03\\
2M00210231+6422070 & 0.06$\pm$0.25 &  &  &  &  &  &  &  \\
2M00213347+6421532 & -0.05$\pm$0.15 & -0.20$\pm$0.06 & 1.91$\pm$0.02 & 3.17$\pm$0.03 & 2.50$\pm$0.03 & -0.13$\pm$0.06 & -0.06$\pm$0.05 & -0.13$\pm$0.04\\
2M00215633+6420081 & -0.19$\pm$0.21 & -0.29$\pm$0.06 & 1.69$\pm$0.02 & 3.64$\pm$0.02 & 2.88$\pm$0.02 & +0.11$\pm$0.06 & +0.17$\pm$0.05 & +0.11$\pm$0.03\\
2M00225591+6427373 & -0.07$\pm$0.36 & -0.69$\pm$0.11 &  &  &  &  &  &  \\ 
2M00232564+6432140 & -0.11$\pm$0.19 & -0.21$\pm$0.06 & 1.60$\pm$0.03 & 3.23$\pm$0.02 & 2.71$\pm$0.03 & +0.10$\pm$0.06 & +0.02$\pm$0.04 & +0.10$\pm$0.03\\
2M00221805+6424390 & -0.09$\pm$0.07 & -0.18$\pm$0.04 & 1.66$\pm$0.01 & 3.80$\pm$0.01 & 2.89$\pm$0.01 & -0.03$\pm$0.06 & +0.02$\pm$0.04 & -0.03$\pm$0.03\\
2M00214423+6425160 & -0.04$\pm$0.06 & -0.18$\pm$0.06 & 1.47$\pm$0.02 & 3.41$\pm$0.02 & 3.21$\pm$0.02 & +0.16$\pm$0.06 & +0.22$\pm$0.05 & +0.16$\pm$0.03\\
2M00213807+6423377 & -0.12$\pm$0.17 & -0.22$\pm$0.05 & 1.67$\pm$0.02 & 3.59$\pm$0.01 & 2.81$\pm$0.02 & +0.05$\pm$0.06 & +0.11$\pm$0.04 & +0.05$\pm$0.03\\
2M00210663+6423385 & -0.15$\pm$0.37 & -0.90$\pm$0.08  &  &  &  &  &  &  \\
2M00215071+6421314 & -0.05$\pm$0.20 & -0.32$\pm$0.06 &1.63$\pm$0.02 & 3.48$\pm$0.02 & 3.08$\pm$0.02 & +0.14$\pm$0.06 & +0.17$\pm$0.04 & +0.14$\pm$0.03\\
2M00213547+6419029 & -0.12$\pm$0.12 & -0.20$\pm$0.04 &1.58$\pm$0.02 & 3.27$\pm$0.03 & 2.82$\pm$0.02 & -0.04$\pm$0.06 & +0.00$\pm$0.05 & -0.05$\pm$0.03\\
2M00215207+6418131 & -0.24$\pm$0.37 &  &  &  &  &  &  &  \\
2M00221438+6418200 & 0.12$\pm$0.18 &  &  &  &  &  &  &  \\ 
2M00224016+6418339 & 0.11$\pm$0.17 &  &  &  &  &  &  &  \\ 
2M00224151+6422225 & -0.05$\pm$0.32 &  &  &  &  &  &  &  \\
2M00222248+6425079 & -0.08$\pm$0.21 & -0.28$\pm$0.07 & 1.57$\pm$0.02 & 3.56$\pm$0.02 & 2.63$\pm$0.02 & +0.00$\pm$0.06 & +0.04$\pm$0.04 & +0.00$\pm$0.03\\
2M00222412+6428578 & -0.77$\pm$0.09 &  &  &  &  &  &  &  \\ 
\hline
\end{tabular}
\end{table*}

The strength of the \mbox{Ca\,{\sc ii}} triplet lines depends not only on the chemical composition 
but also on the temperature and gravity of the star. This dependence can be removed 
because stars of the same chemical composition define a clear 
sequence in the luminosity--$\Sigma$Ca plane when the temperature and/or gravity 
change. In our case we have used as luminosity indicator the absolute magnitude in $V$, $I$, and $K_\mathrm{S}$, 
denoted  $M_V$, $M_I$, and $M_{K_s}$, respectively. The absolute magnitudes of each star were 
obtained using the  extinction coefficients listed in Table 6 from \citet{1998ApJ...500..525S} and using the distance modulus  and  reddening derived in Section~\ref{sectcmd}. 

\begin{figure}
	\includegraphics[width=\columnwidth]{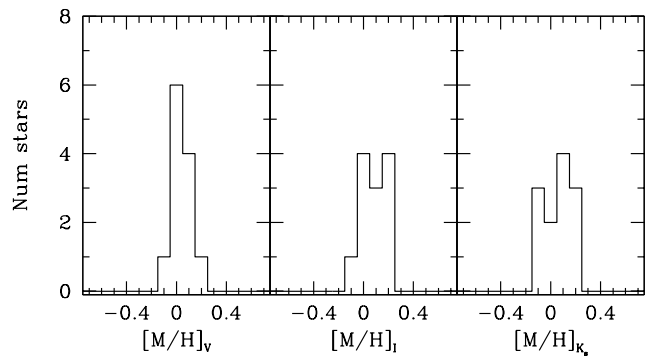}
    \caption{Distribution of the metallicities derived from the strength of \mbox{Ca\,{\sc ii}} of red giant stars located in the innermost 13\arcmin\ and with radial velocities in the range $-53.1\pm2\times3.1$ km s$^{-1}$.}
    \label{fig:catabu}
\end{figure}

The metallicities have been obtained using the relations derived by \citet{2013MNRAS.434.1681C} which are valid for wide ranges in age and metallicity. The obtained values are listed in Table~\ref{tablecat}. The metallicity distribution in each bandpass is shown in Fig.~\ref{fig:catabu}. As before, we have fitted Gaussians to determine the average metallicity in each of them. The obtained values are listed in Table~\ref{catabun}. The column $\left\langle \sigma_{ind}\right\rangle$ represents the average uncertainty in the determination of individual metallicities for each star and bandpass. Finally, the average values of the three bandpasses obtained as the weighted mean are listed in the last row.

Our sample of giants in King~1 includes stars in both red clump and red giant branch. However, the relations used to derive the stellar metallicities were obtained only for red giant branch stars. \citet{2015A&A...578A..27C} demonstrated that the maximum difference in 
$\Sigma$Ca between red clump and red giant branch stars is $\sim$0.16 \AA. 
This implies that the metallicity of a red clump star derived with the red giant branch 
relationships may be over-estimated by a maximum of $\sim$0.07 dex, independent of the 
luminosity indicator used. Although systematic, this value is lower than
the uncertainties of the relationships used and lower than the 
standard deviation of the metallicity determination. There is no calibration for \mbox{Ca\,{\sc ii}} lines for main sequence stars that can be used to determine metallicities for these objects.

\begin{table}
 \centering
\caption{Average metllicity, dispersion and typical error of the individual determinations of each bandpass from \mbox{Ca\,{\sc ii}} lines.\label{catabun}} 
 \begin{tabular}{@{}lcccc@{}}
\hline
Bandpass  & $\left\langle [M/H]\right\rangle$ & $\sigma$ & $\left\langle \sigma_{ind}\right\rangle$\\
   \hline
$V$ & +0.06 & 0.07 & 0.09 \\
$I$ & +0.09 & 0.09 &  0.08 \\
$K_\mathrm{S}$ & +0.06 & 0.09 &  0.09 \\
Average & +0.07$\pm$0.02 & 0.08$\pm$0.01 & 0.09\\
\hline
\end{tabular}
\end{table}

The metallicities derived by fitting the whole spectrum with synthetic spectra for both giants and main sequence stars are in general lower than those obtained from the strengths of the three \mbox{Ca\,{\sc ii}} lines. However, they are still within the uncertainties taken into account that the first method has an average uncertainty of 0.20 dex. Anyway, since the determination of metallicities from the infrared \mbox{Ca\,{\sc ii}} lines have been more widely tested we consider this result, [M/H]=+0.07$\pm$0.08 dex, as our metallicity determination for King~1. This is the value used in the following Section.

\section{King~1 in the Galactic disc context}\label{sectrends}

\begin{figure*}
	\includegraphics[width=\textwidth]{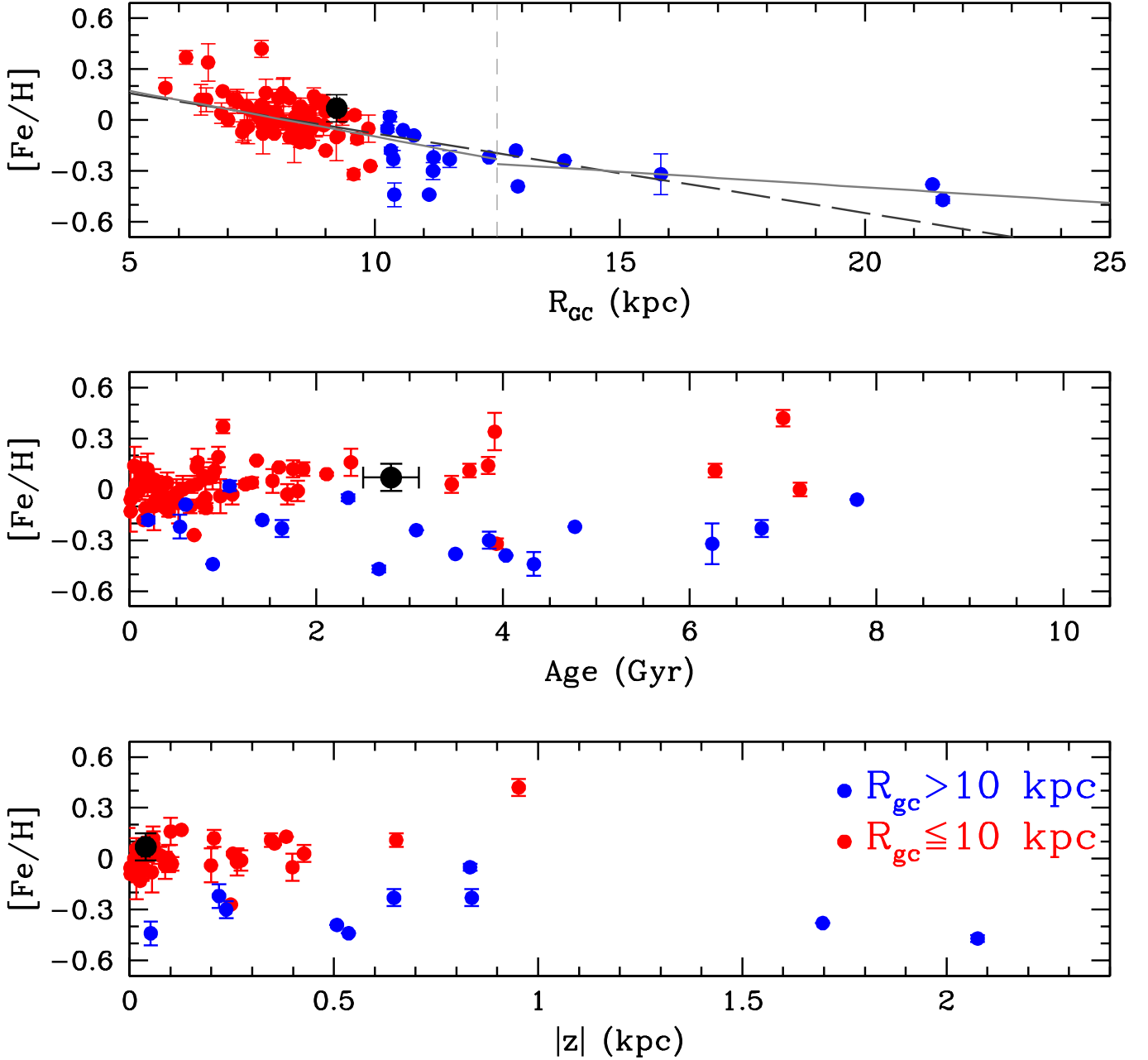}
    \caption{The run of [Fe/H] as a function of $R_{\rm GC}$ (top), age (middle), and $|z|$ (bottom) of the open clusters in the \citet{2016A&A...585A.150N} compilation. Red and blue points are clusters inside and outside galactocentric distances of 10 kpc, respectively. King~1 is represented by  the black point in each panel. Solid lines in the top panel represent two separate linear fits obtained for open clusters inside and outside 12.5 kpc. The dashed line is the fit obtained by the same authors using all open clusters.}
    \label{fig:trends}
\end{figure*}

It is helpful to discuss the behavior of King~1 in comparison with the bulk of open clusters. To do this comparison we used the compilation of clusters obtained by \citet{2016A&A...585A.150N}. Briefly, this sample includes high [Fe/H] abundances available in the literature from high-resolution spectroscopy. We refer the reader to that paper for more details. We compare [Fe/H] abundances derived from iron lines with overall metallicities, [M/H], derived from a calibration of calcium lines. Owing to the way that the values derived from the infrared lines of the \mbox{Ca\,{\sc ii}} we consider that both are in the same scale. In Fig.~\ref{fig:trends} we have plotted the run of metallicities with Galactocentric distance, $R_{\rm GC}$, age, and vertical distance to the Galactic plane, $|z|$, for this sample in the top, middle, and bottom panels, respectively. As a reference, we have over-plotted in the top panel  the linear fit to the whole sample. We show as well the effect of a change in the slope at  $R_{\rm GC}\sim$12.5 kpc. King~1 is represented by the black point in each panel using the values derived here.

In general, King~1 closely follows the trends observed in the Galactic disc. With an age of 2.8 Gyr King~1 is older than the bulk of open clusters. The obtained metallicity for King~1 is slightly high for a cluster located at the Galactocentric distance of King~1, but still within the uncertainties. This may suggest that King~1 has been formed in an inner radius and it has migrated to its current location. For instance, clusters with the same metallicity than King~1 are found at $R_{\rm GC}\sim$6.5 kpc \citep[e.g.][]{2016A&A...591A..37J}. As other open clusters older than 2 Gyr within $R_{\rm GC}<$10 kpc King~1 is more metal-rich than those clusters with older ages located in the outer disc. Finally, King~1 is located almost in the Galactic plane. It  shares a similar metallicity with other systems located at low vertical distances. 

In our analysis we have also estimated the [$\alpha$/M] ratio. The obtained of $\left\langle [\alpha/M]\right\rangle $=-0.10 dex is within the uncertainties, typically 0.25 dex, consistent with the bulk of open clusters with solar metallicities \citep[e.g.][]{2013ApJ...777L...1F}.

Regarding kinematics, we complement our radial velocities with proper motions available in the literature. From the 189 stars in common with the recent Hot Stuff for One Year (HSOY) catalogue \citep{2017arXiv170102629A}, based on PPMXL \citep{2010AJ....139.2440R} and Gaia DR1. 
We have 76 stars with radial velocities compatible with being members of King~1. The mean proper motion with 1-sigma clipping from the central 13\arcmin, the results is $\mu_{\alpha}\cos\delta=$-2.3$\pm$2.1 mas yr$^{-1}$ and $\mu_{\delta}=$-3.9$\pm$2.2 mas yr$^{-1}$ from 32 stars. The mean radial velocity derived in Section~\ref{sec31} yields to radial velocities with respect to the galactic standard of rest (GSR) and the regional standard of rest (RSR) of $178.3\pm 11.6$ and $-4.5\pm11.8$ km s${^1}$, respectively. The values have been computed as in \citet{2016MNRAS.458.3150C}. In comparison with the clusters studied in that paper, the values of King~1 are in good agreement with those of NGC~7789, in spite of the different ages. The $U_s$, $V_s$ and $W_s$ components are $55.84\pm19.33$, $9.09\pm14.86$ and $23.96\pm10.76$ km s${^1}$, respectively. These components are different from the ones of other clusters in Perseus arm and also different from the mean values for the High Mass Stars Forming Regions in the arm as studied by \citet{2014ApJ...783..130R}. No definitive conclusions can be extracted until the accuracy of the proper motions is improved.

\section{Conclusions}\label{sec:conclusions}

We have analysed wide field photometry and medium resolution spectroscopy for 189 giants and main sequence stars in the area of the second quadrant open cluster King~1. Our main results can be summarized as follows:

\begin{itemize}
\item We determined that the centre of the cluster is located at $\alpha_{2000}=00^{\rm h}22^{\rm m}$ and $\delta_{2000}=+64\degr23\arcmin$ with an uncertainty of $\sim$1\arcmin.
\item From the stellar density profile of King~1 we determined a central density of $\rho_{0}=6.5\pm0.2$ star arcmin$^{-2}$ and a core radius of $r_{\rm core}=1\farcm9\pm0\farcm2$.
\item By comparing the observed color-magnitude diagram with those of other well known open clusters and with isochrones we have estimated an age of $2.8\pm0.3$ Gyr, a distance modulus of $(m-M)_{\rm o}=10.6\pm0.1$ mag and a reddening of $E(B-V)=0.80\pm0.05$ mag. 
\item From the analysis of the radial velocity distribution we determined an average velocity for King~1 stars of  $\left\langle V_{r}\right\rangle $= -53.1$\pm$3.1 km s$^{-1}$.
\item We have obtained that the metallicity of King~1 is  [M/H]=+0.07$\pm$0.08 dex from the strengths of the three infrared \mbox{Ca\,{\sc ii}} lines.
\item  By fitting the whole spectrum with synthetic spectra in both main sequence and giant stars we have determined an $\left\langle [\alpha/M]\right\rangle $=-0.10$\pm$0.08 dex with a typical uncertainty of $\pm$0.25 dex. 
\item We have calculated the proper motion of the cluster from UCAC4 proper
motions as $\mu_{\alpha}\cos\delta=$-4.2$\pm$3.1 mas yr$^{-1}$ and $\mu_{\delta}=$0.6$\pm$3.2 mas yr$^{-1}$, and with the recent HSOY catalogue $\mu_{\alpha}\cos\delta=$-3.2$\pm$2.1 mas yr$^{-1}$ and $\mu_{\delta}=$-3.9$\pm$2.0 mas yr$^{-1}$ obtained for those stars assumed as King~1 members from their radial velocities in the inner 13\arcmin.
\item The properties derived for King~1 match the values of other open clusters with similar ages and location in the Galactic plane.
\end{itemize}

\section*{Acknowledgments}

We acknowledge the anonymous referee for his/her thorough revision which has contributed significantly to increase the quality and clarity of this paper. Part of this work has been performed in the framework of L. Rodr\`{\i}guez Espinosa MSc thesis at the University of La Laguna (Spain). This research made use of the WEBDA database, operated at the Department 
of Theoretical Physics and Astrophysics of the Masaryk University, and the 
SIMBAD database, operated at the CDS, 
Strasbourg, France. This publication makes use of data products from the Two 
Micron All Sky Survey, which is a joint project of the University of 
Massachusetts and the Infrared Processing and Analysis Center/California 
Institute of Technology, funded by the National Aeronautics and Space 
Administration and the National Science Foundation. This work has made use of BaSTI web tools.
This work was supported by the MINECO (Spanish Ministry of Economy) - FEDER through grants ESP2016-80079-C2-1-R, ESP2014-55996-C2-1-R, AYA2013-42781P, AYA2014-56795P and MDM-2014-0369 of ICCUB (Unidad de Excelencia 'Mar\'{\i}a de Maeztu').









\bsp	
\label{lastpage}
\end{document}